# Parabolic scaling in overdoped cuprate: a statistical field theory approach

Yong Tao[†]

College of Economics and Management, Southwest University, Chongqing, China

Department of Management, Technology and Economics, ETH Zurich, Switzerland

**Abstract:** Recently, Bozovic *et al.* reported that [Nature **536**, 309-311 (2016)], in the overdoped side of the single-crystal $La_{2-x}Sr_xCuO_4$ (LSCO) films, the transition temperature $T_c$ and zero-temperature superfluid phase stiffness $\rho_s(0)$ will obey a two-class scaling law: $T_c = \gamma \cdot \sqrt{\rho_s(0)}$ for $T_c \leq T_Q$ and $T_c \propto \rho_s(0)$ for $T_c \geq T_M$, where $\gamma = (4.2 \pm 0.5)\ K^{1/2}$, $T_Q \approx 15\ K$, and $T_M \approx 12\ K$. They further pointed out that the parabolic scaling observed in the highly overdoped side indicates a quantum phase transition from a superconductor to a normal metal. In this paper, we propose a quantum partition function (QPF) for zero-temperature Cooper pairs, by which one can effectively distinguish the mean-field and quantum critical behaviors. We theoretically show that the two-class scaling law can be exactly derived by using the QPF, and the theoretical values of $\gamma$, $T_Q$, and $T_M$ are well in accordance with experimental measure values. Our analyses indicate that the linear scaling $T_c \propto \rho_s(0)$ is a mean-field behavior, while the parabolic scaling $T_c = \gamma \cdot \sqrt{\rho_s(0)}$ is a quantum critical behavior.

[†] Correspondence to: taoyingyong@yahoo.com

## 1. Introduction

Over recent decades, with the great advances in cooling technologies, much attention was focused on investigating the behaviors of Cooper pairs near zero temperature. Among all physical quantities, the zero-temperature superfluid phase stiffness $\rho_s(0)$ is a central parameter for describing zero-temperature Cooper pairs, since it can be exactly obtained by measuring magnetic penetration depths of superconducting materials. For copper oxide materials, there has been much interest for seeking the potential correlations between the transition temperature $T_c$ and $\rho_s(0)$. The earliest pattern was referred to as the Uemura relation [1-2] $T_c \propto \rho_s(0)$, which works reasonably well for the underdoped materials. Later, a more universal relation, the Homes' law [3-6] $T_c \propto \rho_s(0)/\sigma_{dc}$ was found to hold regardless of underdoped, optimally doped, and overdoped materials, where $\sigma_{dc}$ denotes the dc conductivity measured at approximately $T_c$. Theoretically, Homes' law has been well known as a mean-field result of the dirty-limit BCS theory [4, 7-8]. Despite these successes, some scholars questioned the validity of Homes' law in highly underdoped and overdoped sides. For example, the relation between $T_c$ and $\rho_s(0)$ was found to be sub-linear in highly underdoped materials [9-12]. Recently, by investigating the overdoped side of the single-crystal $La_{2-x}Sr_xCuO_4$ films, Bozovic *et al.* observed a two-class scaling law [13]:

$$\begin{cases} T_c = \alpha \cdot \rho_s(0) + T_0, & T_c \geq T_M \\ T_c = \gamma \cdot \sqrt{\rho_s(0)}, & T_c \leq T_Q \end{cases}, \qquad (1)$$

where $T_M \approx 12\,K$, $T_Q \approx 15\,K$, $\alpha = 0.37 \pm 0.02$, $T_0 = (7.0 \pm 0.1)\,K$, and $\gamma = (4.2 \pm 0.5)\,K^{1/2}$. The difference between $T_M$ and $T_Q$ implies that the two-class scaling law (1) is non-smoothly linked by linear and parabolic parts.

Equation (1) indicates that a parabolic scaling emerges in the highly overdoped side [13]. Since the two-class scaling law (1) differs significantly from Homes' law, Bozovic *et al.* concluded that their experimental findings are incompatible with the mean-field description [13-15]. The linear part in equation (1) can be derived by using the dirty-limit BCS theory [4, 7-8], and therefore is a mean-field result; however, the parabolic part may hint potential new physics [13]. As a possible evidence, Bozovic *et al.* have observed that, with increased doping ($T_c \to 0$), $La_{2-x}Sr_xCuO_4$ becomes more metallic, and increased doping induces a quantum phase transition from a superconductor to a normal metal [13-15]. This observation indicates that, when $T_c \to 0$, quantum fluctuations may play an important role for inducing the parabolic scaling in equation (1). In this paper, we propose a quantum partition function for describing quantum critical behaviors of zero-temperature Cooper pairs. Based on such a quantum partition function, we will exactly reproduce the two-class scaling law (1). Here, we adopt the natural units $\hbar = c = k_B = 1$, where $\hbar$ denotes the reduced Planck constant, $c$ is the light speed, and $k_B$ is the Boltzmann constant.

## 2. Quantum partition function for zero-temperature Cooper pairs

The free energy density of zero-temperature Cooper pairs can be generally written as [16]:

$$\mathcal{L} = \sigma \cdot |\partial_\tau \phi(\boldsymbol{q},\tau)|^2 + \eta \cdot |\nabla\phi(\boldsymbol{q},\tau)|^2 + \lambda_2 \cdot |\phi(\boldsymbol{q},\tau)|^2 + \lambda_4 \cdot |\phi(\boldsymbol{q},\tau)|^4, \quad (2)$$

where $\phi(\boldsymbol{q},\tau)$ denotes the order parameter of zero-temperature Cooper pairs, and it is a function of space $\boldsymbol{q}$ and imaginary time $\tau$. Here $\tau \in \left[0,\frac{1}{T}\right]$ with the temperature $T$ being 0. $\sigma$, $\eta$, $\lambda_2$ and $\lambda_4$ are phenomenological parameters [16].

If one denotes the zero-temperature superfluid phase stiffness by $|\phi(\boldsymbol{q},\tau)|^2$, then, by applying Gor'kov's Green function method [8] into the BCS theory at $T = 0$ and $T_c \approx 0$, one can obtain [17]:

$$\eta = 1, \quad (3)$$

$$\lambda_2 = \lambda_2(T_c) = -\frac{24\pi^2 m_e}{7\zeta(3)\cdot\varepsilon_F} T_c^2, \quad (4)$$

$$\lambda_4 = \lambda_4\big(T_c, \rho_s(0)\big) = \frac{12\pi^2 m_e}{7\zeta(3)\cdot\varepsilon_F} \cdot \frac{T_c^2}{\rho_s(0)}, \quad (5)$$

where $\rho_s(0) = n_s(0)/4m_e$ and $n_s(0)$ denote zero-temperature superfluid phase stiffness [13] and zero-temperature superfluid density when materials are homogenous, $\zeta(x)$ is the Riemann zeta function, $\varepsilon_F$ is the Fermi energy, and $m_e$ is the rest mass of an electron. The derivation for equations (3)-(5) can be found in Appendix A, where we have clarified why Gor'kov's method holds at $T = 0$.

Equations (3)-(5) are derived by using the BCS theory, which assumes that quantum fluctuations on all size scales are averaged out. Based on such an assumption of the mean-field, $n_s(0)$ is equal to the total number density of electrons in the normal state [8] and hence can be regarded as a constant. This is the standard explanation of the BCS theory. However, later we will observe that $n_s(0)$ changes with $T_c$ as long as quantum fluctuations cannot be averaged out.

Due to equations (3), (4), and (5), $\sigma$ is the unique phenomenological parameter in equation (2). In this paper, we order $\sigma = 1$ so that the free energy density (2) yields an exact relativistic form:

$$\mathcal{L}(T_c) = |\partial_\tau \phi(\boldsymbol{q},\tau)|^2 + |\boldsymbol{\nabla}\phi(\boldsymbol{q},\tau)|^2 + \lambda_2(T_c)\cdot|\phi(\boldsymbol{q},\tau)|^2 + \lambda_4\big(T_c,\rho_s(0)\big)\cdot |\phi(\boldsymbol{q},\tau)|^4. \quad (6)$$

It is easy to observe that the transition temperature $T_c$ in equation (6) plays the role of temperature $T$ in the classical Landau-Ginzburg free energy. Later, we will show that $T_c = 0$ is a potential critical point. To guarantee the self-consistency of equation (6), we need to verify that $|\phi(\boldsymbol{q},\tau)|^2$ is the zero-temperature superfluid phase stiffness. To this end, the free energy density (6) is varied to obtain the field equation of Cooper pairs:

$$\partial_\tau^2\phi(\boldsymbol{q},\tau) + \nabla^2\phi(\boldsymbol{q},\tau) - \lambda_2\phi(\boldsymbol{q},\tau) - 2\lambda_4\cdot|\phi(\boldsymbol{q},\tau)|^2\phi(\boldsymbol{q},\tau) = 0. \quad (7)$$

For homogenous superconductors, equation (7) yields $|\phi(\boldsymbol{q},\tau)|^2 = -\lambda_2/2\lambda_4 = \rho_s(0)$, where equations (4) and (5) have been used. Because $\rho_s(0)$ denotes the zero-temperature superfluid phase stiffness of homogenous materials, $|\phi(\boldsymbol{q},\tau)|^2$ indeed denotes the zero-temperature superfluid phase stiffness. This verifies the self-consistency of the free energy density (6).

Using the free energy density (6), we propose a quantum partition function (QPF) for zero-temperature Cooper pairs as follows:

$$Z(T_c,J,J^*) = \int[\mathcal{D}\phi(\boldsymbol{q},\tau)^*]_\Lambda \int[\mathcal{D}\phi(\boldsymbol{q},\tau)]_\Lambda\, e^{-\int d\tau \int d^D\boldsymbol{q}[\mathcal{L}(T_c) - J(\boldsymbol{q},\tau)\phi(\boldsymbol{q},\tau) - J(\boldsymbol{q},\tau)^*\phi(\boldsymbol{q},\tau)^*]}, \quad (8)$$

where $J(\boldsymbol{q},\tau)$ denotes the external field, $\Lambda$ is the momentum cut-off, and $D$ is the dimension of superconducting materials.

From a perspective of effective field theory, a quantum field theory should be defined fundamentally with a cut-off $\Lambda$ [18-20]. For the crystal materials, a rigid renormalization theory can be defined on a cubic lattice of a lattice unit:

$$a = \frac{1}{\Lambda}, \tag{9}$$

where $a$ denotes the minimal lattice constant. The physical meaning of equation (9) is that quantum fluctuations with wavelengths less than $2\pi a$ can be averaged out [19]. Weinberg also pointed out that [21], in solid-state physics, there really is a cut-off, the lattice spacing $a$, which one must take seriously in dealing with phenomena at similar length scales.

Since the momentum cut-off $\Lambda$ is determined by $a$, there is no longer any phenomenological parameter in the QPF (8). Therefore, the validity of the QPF (8) can be justified by the experimental investigation result (1).

## 3. Parabolic scaling

We assume that quantum fluctuations with wavelengths larger than $2\pi a$ cannot be averaged out. By the theory of critical phenomena, this means that the coefficients $\lambda_2(T_c)$ and $\lambda_4\big(T_c,\rho_s(0)\big)$ in equation (6) should receive the contributions from quantum fluctuations on these size scales. To evaluate the contributions, by applying the renormalization group approach to the QPF (8) one can obtain the renormalization group equations[1] [17]:

$$\frac{d\lambda_2(T_c)}{dlnb} = \lambda_2(T_c)\cdot\big(2-4\hat{\lambda}_4\big) + O\big(\hat{\lambda}_4^2\big), \tag{10}$$

$$\frac{d\hat{\lambda}_4}{dlnb} = (3-D)\cdot\hat{\lambda}_4 - 10\hat{\lambda}_4^2 + O\big(\hat{\lambda}_4^3\big), \tag{11}$$

where the quantum dynamical exponent $z$ is equal to 1 and

[1] $b$ denotes the parameter that guarantees the rescaling transformation $\boldsymbol{q}' = b^{-1}\boldsymbol{q}$ and $\tau' = b^{-z}\tau$, where $z$ is the quantum dynamical exponent [17].

$$\hat{\lambda}_4 = \lambda_4\big(T_c, \rho_s(0)\big) \cdot \frac{(\pi)^{\frac{D}{2}}\Lambda^{D-3}}{2(2\pi)^D\Gamma\left(\frac{D}{2}\right)}. \tag{12}$$

By equations (10)-(12), it is easy to get a nontrivial fixed point:

$$\begin{cases} \lambda_2(T_c) \approx 0 \\ \lambda_4\big(T_c, \rho_s(0)\big) \approx \frac{3-D}{10} \cdot \frac{2(2\pi)^D\Gamma\left(\frac{D}{2}\right)}{(\pi)^{\frac{D}{2}}\Lambda^{D-3}} \end{cases}. \tag{13}$$

$\lambda_2(T_c)$ and $\lambda_4\big(T_c, \rho_s(0)\big)$ are defined by $T_c$ and $\rho_s(0)$ via equation (4) and (5). Substituting equations (4) and (5) into equation (13) yields:

$$\begin{cases} T_c \approx 0 \\ T_c \approx \gamma(D) \cdot \sqrt{\rho_s(0)} \end{cases}, \tag{14}$$

where

$$\gamma(D) = \sqrt{(3-D) \cdot \Lambda^{3-D} \cdot \frac{7(2\pi)^D\Gamma\left(\frac{D}{2}\right)\zeta(3)\cdot\varepsilon_F}{60(\pi)^{\frac{D}{2}+2}m_e}}. \tag{15}$$

If we denote $T_c \approx 0$ by $T_c \le T_Q(D)$, equation (14) can be written in the form:

$$T_c = \gamma(D) \cdot \sqrt{\rho_s(0)} \text{ for } T_c \le T_Q(D), \tag{16}$$

where $T_Q(D)$ denotes a sufficiently low temperature. The physical meaning of equation (16) is that $\rho_s(0)$ will change with $T_c$ as long as $T_c \le T_Q(D)$. Later we will theoretically show $T_Q(2) \le \gamma(2)^2$ and $T_Q(3) \le 0$.

The two-class scaling law (1) was found in the single-crystal $La_{2-x}Sr_xCuO_4$ films ($D = 2$) around $x = 0.25$ [13]. Therefore, for $D = 2$, equation (16) reproduces the parabolic part in the two-class scaling law (1). To verify this, we show that $\gamma(2)$ is in accordance with the existing experimental measure value. Plugging equation (9) into equation (15) one can obtain [22]:

$$\gamma(2) = \sqrt{\frac{7\cdot\zeta(3)\cdot\varepsilon_F}{15\cdot\pi\cdot a\cdot m_e}}. \tag{17}$$

For single-crystal $La_{2-x}Sr_xCuO_4$ films, substituting the data $a \approx 3.8 \times 10^{-10}$ m [13] and $\varepsilon_F(x \approx 0.2) \approx 8.75$ eV [23] into equation (17) yields [22]:

$$\gamma(2) \approx 4.29\, K^{1/2}, \tag{18}$$

which exactly agrees with the experimental value $(4.2 \pm 0.5)\, K^{1/2}$ [13].

The high accordance between theoretical and experimental values thoroughly proves that the parabolic scaling in equation (1) is due to quantum fluctuations. From this meaning, the nontrivial fixed point (13) describes the quantum critical behaviors of zero-temperature Cooper pairs when $T_c \leq T_Q(D)$. However, we do not clarify the range of applicability of the nontrivial fixed point (13), i.e., the value of $T_Q(D)$. According to the renormalization group theory, the nontrivial fixed point (13) is valid if and only if quantum fluctuations cannot be averaged out. Therefore, to evaluate $T_Q(D)$, we need to find a criterion for identifying the validity of the mean-field approximation.

## 4. Quantum Ginzburg number

For thermal fluctuations, there exists a clear criterion of the applicability of the mean-field theory, i.e., the classical Ginzburg number $G_i$ [24-26], where the mean-field approximation is valid when $G_i \ll 1$. To evaluate quantum fluctuations, we extend $G_i$ to a quantum version. To this end, let us first define the correlation function of the order parameter $\phi(\boldsymbol{q},\tau)$ as [16]:

$$G(\boldsymbol{q}-\boldsymbol{q}',\tau-\tau') = \langle[\phi(\boldsymbol{q},\tau)-\langle\phi(\boldsymbol{q},\tau)\rangle]\cdot[\phi(\boldsymbol{q}',\tau')^*-\langle\phi(\boldsymbol{q}',\tau')^*\rangle]\rangle, \quad (19)$$

where the mean-value of a physical variable $A(\boldsymbol{q},\tau)$ is defined by

$$\langle A(\boldsymbol{q},\tau)\rangle = \frac{1}{Z(T_c,J,J^*)}\int \mathcal{D}\phi(\boldsymbol{q},\tau)^*\int \mathcal{D}\phi(\boldsymbol{q},\tau)\, e^{-\int d\tau \int d^D\boldsymbol{q}[\mathcal{L}(T_c)-J(\boldsymbol{q},\tau)\phi(\boldsymbol{q},\tau)-J(\boldsymbol{q},\tau)^*\phi(\boldsymbol{q},\tau)^*]}A(\boldsymbol{q},\tau). \quad (20)$$

Using equations (8), (19) and (20), it is easy to obtain:

$$G(\boldsymbol{q}-\boldsymbol{q}',\tau-\tau') = \frac{\partial^2 lnZ(T_c,J,J^*)}{\partial J(\boldsymbol{q},\tau)\partial J(\boldsymbol{q}',\tau')^*} = \frac{\partial\langle\phi(\boldsymbol{q},\tau)\rangle}{\partial J(\boldsymbol{q}',\tau')^*}. \quad (21)$$

As a quantum extension of the classical Ginzburg number $G_i$, by using the correlation function (19) we construct an error function of the order parameter $\phi(\boldsymbol{q},\tau)$ as follows:

$$e^q(D) = \frac{\left|\int_0^\infty d\tau \int d^D\boldsymbol{q} G(\boldsymbol{q},\tau)\right|}{\int_0^\infty d\tau \int d^D\boldsymbol{q}\phi(\boldsymbol{q},\tau)^*\phi(\boldsymbol{q},\tau)}, \quad (22)$$

where $e^q(D)$ returns to the classical Ginzburg number $G_i$ when $\phi(\boldsymbol{q},\tau)$ is independent of $\tau$; that is, $e^q(D) = |\int d^D\boldsymbol{q}G(\boldsymbol{q})|/\int d^D\boldsymbol{q}\,\phi(\boldsymbol{q})^*\phi(\boldsymbol{q}) = G_i$ if $\phi(\boldsymbol{q},\tau)=\phi(\boldsymbol{q})$. By equation (22), the mean-field approximation is valid if and only if

$$e^{q}(D)\ll 1. \tag{23}$$

Therefore, when the inequality (23) breaks down, the nontrivial fixed point (13) holds. To rigidly determine the range of applicability of the nontrivial fixed point (13), we need to explore the physical meaning of the inequality (23). To this end, let us order

$$M(T_c)=\left|\int_0^{\infty}d\tau\int d^D\boldsymbol{q}G(\boldsymbol{q},\tau)\right|, \tag{24}$$

$$W(t)=\int_0^{\frac{1}{t}}d\tau\int d^D\boldsymbol{q}\,\phi(\boldsymbol{q},\tau)^{*}\phi(\boldsymbol{q},\tau). \tag{25}$$

By using equations (24) and (25), equation (22) can be written as $e^{q}(D)=M(T_c)/W(0)$. Obviously, we have $W(t)\le W(0)$ and $M(T_c)\le W(0)$. Since $G(\boldsymbol{q},\tau)$ is the correlation function, $M(T_c)$ actually denotes the magnitude of quantum fluctuations. Thus, the physical meaning of the inequality (23) is that quantum fluctuations can be omitted if and only if their magnitude is extremely small; that is, $M(T_c)\ll W(0)$. Based on this observation, there should exist a critical magnitude $M_0$ so that when $M(T_c)\ge M_0$, quantum fluctuations cannot be omitted. This means that the nontrivial fixed point (13) is valid when $M(T_c)\ge M_0$. To evaluate the value of $M(T_c)$, we introduce an approximation $\phi(\boldsymbol{q},\tau)\approx\langle\phi(\boldsymbol{q},\tau)\rangle\approx\langle\phi(\boldsymbol{q},\tau)\rangle_{vac}$. This approximation has been well known for evaluating the magnitude of thermal fluctuations when $T>0$ [24-25].

**Proposition 1:** If $\phi(\boldsymbol{q},\tau)\approx\langle\phi(\boldsymbol{q},\tau)\rangle\approx\langle\phi(\boldsymbol{q},\tau)\rangle_{vac}$, then the magnitude of quantum fluctuations, $M(T_c)$, yields:

$$M(T_c)=\xi^2\propto T_c^{-2}, \tag{26}$$

where $\xi=\left(-\lambda_2(T_c)\right)^{-1/2}$ denotes the quantum correlation length[2] and $\langle\phi(\boldsymbol{q},\tau)\rangle_{vac}$ denotes the vacuum expectation value of $\langle\phi(\boldsymbol{q},\tau)\rangle$.

***Proof.*** See Appendix B. □

---

[2] Equation (26) implies $\xi\propto T_c^{-\delta}$ with a critical exponent $\delta$ being 1. If we consider the two-order correction from the renormalization group, the quantum critical exponent $\delta$ for $D=2$ should yield 1.25. This is a new prediction that can be tested. We propose that one can measure $\delta$ by using neutron scattering experiments near $T_c=0$, which have been successfully carried out for measuring the critical exponent of the thermal correlation length [27].

By equation (26), the magnitude $M(T_c)$ and the correlation length $\xi$ grow as $T_c$ decreases, and both of them finally diverge at $T_c = 0$. This implies that $T_c = 0$ is a critical point. Since $M(T_c)$ increases as $T_c$ declines, there does exist $T_Q'$ so that when $T_c \le T_Q'$, one has $M(T_c) \ge M_0$. This means that the nontrivial fixed point (13) is valid when $T_c \le T_Q'$. To estimate $T_Q'$, we construct an index as below:

$$E^q(D,t) = \frac{M(T_c)}{W(t)}. \tag{27}$$

It is easy to check $E^q(D,0) = e^q(D)$ and $E^q(D,t) \ge 0$. If we order $W(T^*) = M_0$, then $E^q(D,T^*) \ge 1$ is equivalent to $M(T_c) \ge M_0$, where $W(t) \le W(0)$ and $M(T_c) \le W(0)$ have been used. Thus, the following proposition provides a way for estimating $T_Q'$.

**Proposition 2:** Let us order $T_Q = T_Q(D) = min\{T^*, T_Q'\}$. $E^q(D,T_Q) \ge 1$ leads to $E^q(D,T^*) \ge 1$.

***Proof.*** Since $T_Q \le T^*$, we have $W(T_Q) \ge W(T^*)$, which leads to $E^q(D,T_Q) = M(T_c)/W(T_Q) \le M(T_c)/W(T^*) = E^q(D,T^*)$. That is to say, $E^q(D,T_Q) \ge 1$ leads to $E^q(D,T^*) \ge 1$. □

Since $E^q(D,T^*) \ge 1$ is equivalent to $M(T_c) \ge M_0$, by the Proposition 2 $E^q(D,T_Q) \ge 1$ leads to $M(T_c) \ge M_0$. Therefore, we conclude that the nontrivial fixed point (13) is valid when $E^q(D,T_Q) \ge 1$. Since $T_Q$ is the lower bound of $T_Q'$ and the nontrivial fixed point (13) is equivalent to equation (16), we have the following criterion:

**Criterion A:** If $E^q(D,T_Q) \ge 1$, the parabolic scaling (16) holds for $T_c \le T_Q$.

To estimate $T_Q$ by using the Criterion A, we need to calculate $E^q(D,T_Q)$. Since $M(T_c)$ has been estimated by equation (26), we only calculate the value of $W(T_Q)$. As an approximation, we consider that the integral scope of $\int d^D\boldsymbol{q}\, \phi(\boldsymbol{q},\tau)^*\phi(\boldsymbol{q},\tau)$ is up to the correlation length $\xi$. Thus, by using $\phi(\boldsymbol{q},\tau) \approx \langle\phi(\boldsymbol{q},\tau)\rangle_{vac}$, we have:

$$W(T_Q) \approx \frac{1}{T_Q}\xi^D |\langle\phi(\boldsymbol{q},\tau)\rangle_{vac}|^2 = \frac{1}{T_Q}\xi^D \rho_s(0). \quad (28)$$

Substituting equations (26) and (28) into $E^q(D,T_Q)$ yields:

$$E^q(D,T_Q) = \frac{T_Q\xi^{2-D}}{\rho_s(0)}. \quad (29)$$

We now estimate $T_Q(D)$ by using equation (29). The Criterion A indicates that $T_c = \gamma(D)\cdot\sqrt{\rho_s(0)}$ holds at $T_c = T_Q(D)$; that is, $T_Q(D) = \gamma(D)\cdot\sqrt{\rho_s(0)}$. Substituting it into $E^q(D,T_Q) \geq 1$ obtains $E^q(D,T_Q) = \xi^{2-D}\gamma(D)^2/T_Q(D) \geq 1$, which indicates:

$$T_Q(D) \leq \xi^{2-D}\gamma(D)^2. \quad (30)$$

For $D = 2$, the inequality (30) yields:

$$T_Q(2) \leq \gamma(2)^2, \quad (31)$$

which by using the experimental value $\gamma(2) \approx 4.2\,K^{1/2}$ yields $T_Q(2) \leq 17\,K$, agreeing with the experimental measure value $T_Q(2) \approx 15\,K$ [13].

For $D = 3$, substituting $\gamma(3) = 0$ into the inequality (30) obtains

$$T_Q(3) \leq 0, \quad (32)$$

which indicates that the parabolic scaling (16) holds for $T_c \leq T_Q(3) = 0$. That is to say, the mean-field approximation always holds for $D = 3$. In fact, Tao has pointed out [17] that $D = 3$ is the upper critical dimension of quantum critical systems, and that the mean-field approximation is valid at the upper critical dimension. Therefore, our result for $D = 3$ agrees with the previous analysis [17].

## 5. The two-class scaling

By using Abrikosov-Gor'kov's mean-field theory for superconducting alloys, for dirty BCS superconductors the relation between $T_c$ and $\rho_s(0)$ can be derived as [7-8, 17, 28]:

$$T_c = \alpha\cdot\rho_s(0) + T_0. \quad (33)$$

The derivation for equation (33) can be found in Appendix C. In particular, by using the latest experimental data [29], Khodel et al [28] have produced the correct theoretical value of $\alpha$. This is an evidence for supporting the linear scaling in equation (1) as a result of Abrikosov-Gor'kov's mean-field theory. By equation (1), equation (33) holds for $T_c \geq T_M$. By the Criterion A, if the mean-field approximation is valid, $E^q(D, T_Q) \leq 1$ should hold. Using equation (27) and $T_M \leq T_Q$, it is easy to verify $E^q(D, T_M) \leq E^q(D, T_Q)$. This implies that one can estimate $T_M$ by using $E^q(D, T_M) \leq 1$. The following proposition will rigidly confirm this fact.

**Proposition 3:** Let us order $\Omega = \int d^D \boldsymbol{q}\, \phi(\boldsymbol{q}, \tau)^* \phi(\boldsymbol{q}, \tau)$. If $\partial\Omega/\partial\tau = 0$ and $T_M > 0$, then we have:

$$e^q(D) \ll E^q(D, T_M). \qquad (34)$$

***Proof.*** See Appendix D. □

**Corollary 1:** If $E^q(D, T_M) \leq 1$, then we have $e^q(D) \ll 1$.

Regarding the Proposition 3, the condition $\partial\Omega/\partial\tau = 0$ should approximately hold as long as $\phi(\boldsymbol{q}, \tau) \approx \langle \phi(\boldsymbol{q}, \tau) \rangle_{vac}$ is satisfied. Thus, by the Corollary 1, we can replace $e^q(D) \ll 1$ by $E^q(D, T_M) \leq 1$ to estimate $T_M$. Since superconducting films imply $D = 2$, by equation (29) we have $E^q(2, T_M) = T_M/\rho_s(0)$. By equation (1), $T_c = \alpha \cdot \rho_s(0) + T_0$ holds at $T_c = T_M$. Substituting $T_M = \alpha \cdot \rho_s(0) + T_0$ into $E^q(2, T_M) \leq 1$ yields $E^q(2, T_M) = \alpha T_M/(T_M - T_0) \leq 1$, indicating

$$T_M \geq \frac{T_0}{1-\alpha}, \qquad (35)$$

where we have considered $0 < \alpha < 1$ [13] and $\rho_s(0) \geq 0$.

Substituting experimental data $\alpha \approx 0.37$ and $T_0 \approx 7K$ into the inequality (35) obtains $T_M \geq 11K$, which agrees with the experimental value $T_M \approx 12K$ [13].

Using equations (16), (31), (33) and (35), we exactly produce the two-class scaling law for $D = 2$ as below:

$$\begin{cases} T_c = \alpha \cdot \rho_s(0) + T_0, & T_c \geq T_M \approx \frac{T_0}{1-\alpha} \\ T_c = \gamma(2) \cdot \sqrt{\rho_s(0)}, & T_c \leq T_Q \approx \gamma(2)^2 \end{cases}, \qquad (36)$$

where $\gamma(2)=\sqrt{\frac{7\cdot\zeta(3)\cdot\varepsilon_F}{15\cdot\pi\cdot a\cdot m_e}}$.

[Insert Table 1 here]

The theoretical values of $\gamma(2)$, $T_Q$, and $T_M$ have been listed in the Table 1. They agree with experimental measure values. In particular, the difference between $T_M\approx 11\ K$ and $T_Q\approx 17\ K$ implies that the part over $[T_M,T_Q]$ should be a combination of linear and parabolic scaling. Here we have fitted equation (36) to experimental data in the Figure 1. The accordance between theoretical formula and experimental data is pretty well. Equation (36) is the main result of this paper. It can be rigidly tested by investigating other quasi-two-dimensional BCS-like superconductors.

[Insert Figure 1 here]

## 6. Conclusion

In conclusion, by using the BCS theory, we propose a quantum partition function (QPF) to describe quantum critical behaviors of zero-temperature Cooper pairs. It was recently found that, in the overdoped side of the single-crystal $La_{2-x}Sr_xCuO_4$ films, a two-class scaling law emerges as: $T_c=\gamma\cdot\sqrt{\rho_s(0)}$ for $T_c\le T_Q$ and $T_c=\alpha\cdot\rho_s(0)+T_0$ for $T_c\ge T_M$. By using the QPF, we show that the parabolic scaling $T_c=\gamma\cdot\sqrt{\rho_s(0)}$ can be exactly derived when $T_c$ is sufficiently low, where the theoretical value of $\gamma$ is exactly calculated as $4.29\ K^{1/2}$, being in accordance with the experimental measure value $\gamma=(4.2\pm0.5)\ K^{1/2}$. Furthermore, we show that the linear scaling $T_c=\alpha\cdot\rho_s(0)+T_0$ is a mean-field behavior of the dirty-limit BCS theory, which lies far beyond the control of the QPF. To determine the range of applicability of the QPF, we extend the classical Ginzburg number to a quantum version. By using the quantum Ginzburg number, we show that the QPF holds for $T_c\le T_Q$, while the mean-field theory holds for $T_c \geq T_M$, where theoretical values of $T_Q$ and $T_M$ are estimated as $T_Q \approx 17\ K$ and $T_M \approx 11\ K$, respectively, agreeing with experimental measure values $15\ K$ and $12\ K$. The high accordance of theoretical values of $\gamma$, $T_Q$, and $T_M$ with experimental measure results justifies the validity of the QPF. Finally, the QPF predicts that, for 2-dimensional overdoped cuprate films, the transition temperature $T_c$ and the quantum correlation length $\xi$ will obey a scaling $\xi \propto T_c^{-\delta}$ with a critical exponent $\delta$ being around 1.25. This is a new prediction that can be tested. We propose that one can measure $\delta$ by using neutron scattering experiments near $T_c = 0$, which have been successfully carried out for measuring the critical exponent of the thermal correlation length [27].

# Appendices

## A. Derivation for $\boldsymbol{\eta}$, $\boldsymbol{\lambda_2}$, and $\boldsymbol{\lambda_4}$

By using the BCS Hamiltonian of superconductivity, Gor'kov has shown that, when $|T - T_c| \approx 0$, the Landau-Ginzburg equation can be written in the form [8]:

$$\frac{1}{4m_e^*}\nabla^2\psi(T) - \frac{1}{\lambda}\cdot\frac{(T-T_c)}{T_c}\psi(T) - \frac{1}{\lambda\cdot n_s(0)}|\psi(T)|^2\psi(T) = 0, \tag{A.1}$$

where $\lambda = \frac{7\zeta(3)\cdot\varepsilon_F}{6\pi^2 T_c^2}$ and $|\psi(T)|^2$ denotes the superfluid density at the temperature $T$. Moreover, $n_s(0)$ denotes the zero-temperature superfluid density when materials are homogenous, $\zeta(x)$ is the Riemann zeta function, $\varepsilon_F$ is the Fermi energy, and $m_e^*$ is the mass of an electron. Quantitatively, $n_s(0)$ is equal to the total number density of electrons in the normal state [8]. This is the standard description of the BCS theory.

We first verify that Gor'kov's equation (A.1) holds at $T = 0$. Since $|\psi(T)|^2$ denotes the superfluid density at the temperature $T$, we should conclude, for homogenous materials, $|\psi(0)|^2 = n_s(0)$ as long as Gor'kov's equation (A.1) holds at $T = 0$. That is to say, when $|\psi(0)|^2 = n_s(0)$, the self-consistency of equation (A.1) at $T = 0$ can be justified.

When materials are homogenous, $\psi(T)$ is independent of the space $\boldsymbol{q}$. Then, equation (A.1) yields:

$$\frac{1}{\lambda}\cdot\frac{(T-T_c)}{T_c}\psi(T) + \frac{1}{\lambda\cdot n_s(0)}|\psi(T)|^2\psi(T) = 0, \tag{A.2}$$

which can be rewritten as

$$|\psi(T)|^2 = n_s(0)\cdot\left(\frac{T_c-T}{T_c}\right). \tag{A.3}$$

By equation (A.3), we obviously have $|\psi(0)|^2 = n_s(0)$. This verifies the self-consistency of equation (A.1) at $T = 0$.

Now we start to derive $\eta$, $\lambda_2$, and $\lambda_4$ in equation (2). By rescaling $\psi(T)$ according to $\phi(T) = \frac{1}{\sqrt{4m_e^*}}\psi(T)$, equation (A.1) yields the following Lagrangian function:

$$\mathcal{L}(T) = |\nabla\phi(T)|^2 + \frac{4m_e^*}{\lambda} \cdot \frac{(T-T_c)}{T_c} \cdot |\phi(T)|^2 + \frac{8m_e^{*2}}{\lambda \cdot n_s(0)} \cdot |\phi(T)|^4. \tag{A.4}$$

If we order $\rho_s(T) = |\phi(T)|^2$, then $\rho_s(T) = \frac{|\psi(T)|^2}{4m_e^*}$ denotes the superfluid phase stiffness at the temperature $T$. Thus, by equation (A.3), we have:

$$\rho_s(0) = \frac{n_s(0)}{4m_e^*}. \tag{A.5}$$

Substituting equation (A.5) into equation (A.4) yields:

$$\mathcal{L}(T) = |\nabla\phi(T)|^2 + \frac{24\pi^2 m_e}{7\zeta(3)\cdot\varepsilon_F} T_c^2 \cdot \frac{(T-T_c)}{T_c} \cdot |\phi(T)|^2 + \frac{12\pi^2 m_e}{7\zeta(3)\cdot\varepsilon_F} \cdot \frac{T_c^2}{\rho_s(0)} \cdot |\phi(T)|^4. \tag{A.6}$$

If we introduce the imaginary time $\tau \in \left[0, \frac{1}{T}\right]$ with $T = 0$, then we have $\phi(\boldsymbol{q}, \tau) = \phi(0)$ [16]. Since equation (A.1) holds at $T = 0$, we conclude that equation (A.6) holds at $T = 0$ as well. Therefore, by equation (A.6) we have:

$$\mathcal{L}(0) = |\nabla\phi(\boldsymbol{q}, \tau)|^2 - \frac{24\pi^2 m_e}{7\zeta(3)\cdot\varepsilon_F} T_c^2 \cdot |\phi(\boldsymbol{q}, \tau)|^2 + \frac{12\pi^2 m_e}{7\zeta(3)\cdot\varepsilon_F} \cdot \frac{T_c^2}{\rho_s(0)} \cdot |\phi(\boldsymbol{q}, \tau)|^4, \tag{A.7}$$

where we assume $m_e^* = m_e$ at $T = 0$ and $m_e$ denotes the rest mass of an electron.

Comparing equations (2) and (A.7) we have:

$$\eta = 1, \tag{A.8}$$

$$\lambda_2 = \lambda_2(T_c) = -\frac{24\pi^2 m_e}{7\zeta(3)\cdot\varepsilon_F} T_c^2, \tag{A.9}$$

$$\lambda_4 = \lambda_4\big(T_c, \rho_s(0)\big) = \frac{12\pi^2 m_e}{7\zeta(3)\cdot\varepsilon_F} \cdot \frac{T_c^2}{\rho_s(0)}. \tag{A.10}$$

## B. Proof of Proposition 1

***Proof.*** By equation (8), it is easy to obtain the field equation of zero-temperature Cooper pairs as below:

$$[\partial_\tau^2 + \nabla^2 - \lambda_2 - 2\lambda_4 \cdot |\phi(\boldsymbol{q}, \tau)|^2] \cdot \phi(\boldsymbol{q}, \tau) = -J(\boldsymbol{q}, \tau)^*. \tag{B.1}$$

Substituting $\phi(\boldsymbol{q}, \tau) \approx \langle\phi(\boldsymbol{q}, \tau)\rangle$ into equation (B.1) yields:

$$[\partial_\tau^2 + \nabla^2 - \lambda_2 - 2\lambda_4 \cdot |\langle\phi(\boldsymbol{q}, \tau)\rangle|^2] \cdot \langle\phi(\boldsymbol{q}, \tau)\rangle = -J(\boldsymbol{q}, \tau)^*. \tag{B.2}$$

Using equation (21), equation (B.2) can be written in the form:

$$[\partial_\tau^2 + \nabla^2 - \lambda_2 - 4\lambda_4 \cdot |\langle\phi(\boldsymbol{q}, \tau)\rangle|^2] \cdot G(\boldsymbol{q} - \boldsymbol{q}', \tau - \tau') = -\delta(\boldsymbol{q} - \boldsymbol{q}', \tau - \tau'), \tag{B.3}$$

where $\delta(\boldsymbol{q}, \tau)$ denotes the Dirac function.

By using $\langle\phi(\boldsymbol{q}, \tau)\rangle \approx \langle\phi(\boldsymbol{q}, \tau)\rangle_{vac}$ and equation (6) we have:

$$|\langle\phi(\boldsymbol{q},\tau)\rangle|^2 \approx |\langle\phi(\boldsymbol{q},\tau)\rangle_{vac}|^2 = -\frac{\lambda_2}{2\lambda_4}. \tag{B.4}$$

Substituting equation (B.4) into equation (B.3) obtains

$$[\partial_\tau^2 + \nabla^2 + \lambda_2]\cdot G(\boldsymbol{q},\tau) = -\delta(\boldsymbol{q},\tau). \tag{B.5}$$

Let us consider the Fourier transforms as follows:

$$G(\boldsymbol{q},\tau) = \int_0^\infty \frac{d\omega}{2\pi}\int \frac{d^D\boldsymbol{k}}{(2\pi)^D} e^{i\boldsymbol{k}\cdot\boldsymbol{q}+i\omega\tau}\tilde{G}(\boldsymbol{k},\omega), \tag{B.6}$$

$$\tilde{G}(\boldsymbol{k},\omega) = \int_0^\infty d\tau \int d^D\boldsymbol{q}\, e^{-i\boldsymbol{k}\cdot\boldsymbol{q}-i\omega\tau}G(\boldsymbol{q},\tau), \tag{B.7}$$

$$\delta(\boldsymbol{q},\tau) = \int_0^\infty \frac{d\omega}{2\pi}\int \frac{d^D\boldsymbol{k}}{(2\pi)^D} e^{i\boldsymbol{k}\cdot\boldsymbol{q}+i\omega\tau}. \tag{B.8}$$

Substituting equations (B.6)-(B.8) into equation (B.5) obtains:

$$\tilde{G}(\boldsymbol{k},\omega) = \frac{1}{|\boldsymbol{k}|^2+\omega^2-\lambda_2}. \tag{B.9}$$

Substituting equation (B.9) into equation (B.6) yields:

$$G(\boldsymbol{q},\tau) = \int_0^\infty \frac{d\omega}{2\pi}\int \frac{d^D\boldsymbol{q}}{(2\pi)^D} \frac{e^{i\boldsymbol{k}\cdot\boldsymbol{q}+i\omega\tau}}{|\boldsymbol{k}|^2+\omega^2-\lambda_2} \propto e^{-\frac{|\boldsymbol{q}|}{\xi}}, \tag{B.10}$$

where $\xi = (-\lambda_2)^{-1/2}$ denotes the correlation length.

Using equations (B.7) and (B.9), it is easy to find:

$$\tilde{G}(\boldsymbol{0},0) = \int_0^\infty d\tau \int d^D\boldsymbol{q}\, G(\boldsymbol{q},\tau) = (-\lambda_2)^{-1} = \xi^2. \quad \square$$

## C. Derivation for equation (33)

For isotropic BCS superconductors, by using Abrikosov-Gor'kov's mean-field theory for superconducting alloys one can obtain [17]:

$$\lambda_p^{-2}(0) = \frac{4\pi n_s(0)e^2}{m_e^*}\Delta(0)^2\int_0^\infty \frac{1}{(u^2+\Delta(0)^2)\left(\sqrt{u^2+\Delta(0)^2}+\frac{1}{2\tau_S}\right)}du, \tag{C.1}$$

where $\lambda_p(0)$ denotes the penetration depth at zero temperature, $\tau_s$ denotes the scattering relaxation time, $\Delta(0)$ denotes the energy gap at zero temperature, and $e$ denotes the electron charge.

If we order $y = \frac{u}{\Delta(0)}$, equation (C.1) can be rewritten in the form:

$$\lambda_p^{-2}(0) = \frac{4\pi n_s(0)e^2}{m_e^*}\int_0^\infty \frac{1}{(1+y^2)\left(\sqrt{1+y^2}+\frac{1}{2\tau_S\Delta(0)}\right)}dy. \tag{C.2}$$

We investigate equation (C.2) in terms of two cases; that is, clean and dirty superconductors.

For clean superconductors, we should have $\tau_s \to \infty$; thus, equation (C.2) yields:

$$\lambda_p^{-2}(0) = \frac{4\pi n_s(0)e^2}{m_e^*}\int_0^{\infty}\frac{1}{(1+y^2)^{\frac{3}{2}}}dy = \frac{4\pi n_s(0)e^2}{m_e^*};$$

that is,

$$\lambda_p(0) = \sqrt{\frac{m_e^*}{4\pi n_s(0)e^2}}, \tag{C.3}$$

which is the famous London penetration depth [8].

For dirty superconductors, we simply consider $\tau_s \to 0$; thus, equation (C.2) yields:

$$\lambda_p^{-2}(0) = \frac{8\pi n_s(0)e^2}{m_e^*}\tau_s\Delta(0)\int_0^{\infty}\frac{1}{(1+y^2)}dy + o(\tau_s^2). \tag{C.4}$$

Since $\rho_s(0) \propto \lambda_p^{-2}(0)$ and $\Delta(0) \propto T_c$, equation (C.4) implies:

$$\rho_s(0) \propto T_c, \tag{C.5}$$

which can be generally written as:

$$T_c = \alpha \cdot \rho_s(0) + T_0. \tag{C.6}$$

**D. Proof of Proposition 3**

***Proof***. The following equation obviously holds:

$$\int_0^{\frac{1}{T}} d\tau \int d^D\boldsymbol{q}\, \phi(\boldsymbol{q},\tau)^*\phi(\boldsymbol{q},\tau) = \int_0^{\frac{1}{T_M}} d\tau \int d^D\boldsymbol{q}\, \phi(\boldsymbol{q},\tau)^*\phi(\boldsymbol{q},\tau) + \int_{\frac{1}{T_M}}^{\frac{1}{T}} d\tau \int d^D\boldsymbol{q}\, \phi(\boldsymbol{q},\tau)^*\phi(\boldsymbol{q},\tau). \tag{D.1}$$

Substituting $\Omega = \int d^D\boldsymbol{q}\, \phi(\boldsymbol{q},\tau)^*\phi(\boldsymbol{q},\tau)$ into equation (D.1) and by using $\frac{\partial\Omega}{\partial\tau} = 0$, we obtain:

$$\lim_{T\to 0}\int_0^{\frac{1}{T}} d\tau \int d^D\boldsymbol{q}\, \phi(\boldsymbol{q},\tau)^*\phi(\boldsymbol{q},\tau) = \frac{1}{T_M}\Omega + \lim_{T\to 0}\left(\frac{1}{T} - \frac{1}{T_M}\right)\cdot\Omega. \tag{D.2}$$

Since $\frac{1}{T_M}\Omega \ll \lim_{T\to 0}\left(\frac{1}{T} - \frac{1}{T_M}\right)\cdot\Omega$, by using equation (D.2) we have:

$$\frac{1}{T_M}\Omega \ll \lim_{T\to 0}\left(\frac{1}{T} - \frac{1}{T_M}\right)\cdot\Omega + \frac{1}{T_M}\Omega = \lim_{T\to 0}\int_0^{\frac{1}{T}} d\tau \int d^D\boldsymbol{q}\, \phi(\boldsymbol{q},\tau)^*\phi(\boldsymbol{q},\tau), \tag{D.3}$$

which leads to:

$$\int_0^{\frac{1}{T_M}} d\tau \int d^D\boldsymbol{q}\, \phi(\boldsymbol{q},\tau)^*\phi(\boldsymbol{q},\tau) \ll \lim_{T\to 0}\int_0^{\frac{1}{T}} d\tau \int d^D\boldsymbol{q}\, \phi(\boldsymbol{q},\tau)^*\phi(\boldsymbol{q},\tau). \tag{D.4}$$

By using the inequality (D.4), it is easy to verify:

$$e^q(D) = \lim_{T\to 0} \frac{\left| \int_0^{\frac{1}{T}} d\tau \int d^D \boldsymbol{q} G(\boldsymbol{q},\tau) \right|}{\int_0^{\frac{1}{T}} d\tau \int d^D \boldsymbol{q} \phi(\boldsymbol{q},\tau)^* \phi(\boldsymbol{q},\tau)} \ll \lim_{T\to 0} \frac{\left| \int_0^{\frac{1}{T}} d\tau \int d^D \boldsymbol{q} G(\boldsymbol{q},\tau) \right|}{\int_0^{\frac{1}{T_M}} d\tau \int d^D \boldsymbol{q} \phi(\boldsymbol{q},\tau)^* \phi(\boldsymbol{q},\tau)} = E^q(D, T_M). \ \square$$

**Table 1.** Comparison of theoretical results with experimental measure values [13].

| Parameter | Experimental value | Theoretical value |
|---|---|---|
| $\boldsymbol{\gamma(2)}$ | $(4.2 \pm 0.5)\ K^{1/2}$ | $4.29\ K^{1/2}$ |
| $\boldsymbol{T_Q}$ | $15\ K$ | $17\ K$ |
| $\boldsymbol{T_M}$ | $12\ K$ | $11\ K$ |

**a**

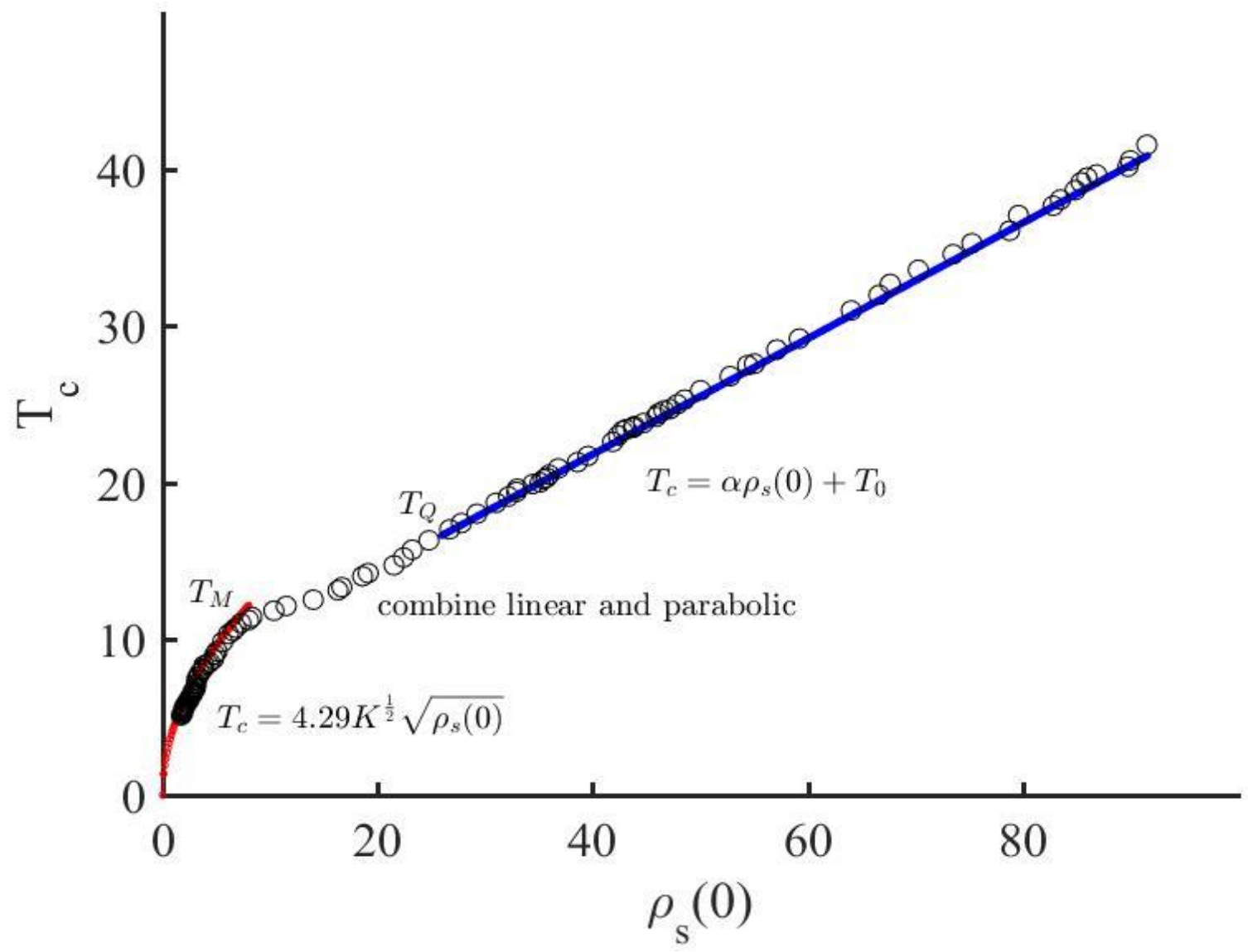

**b**

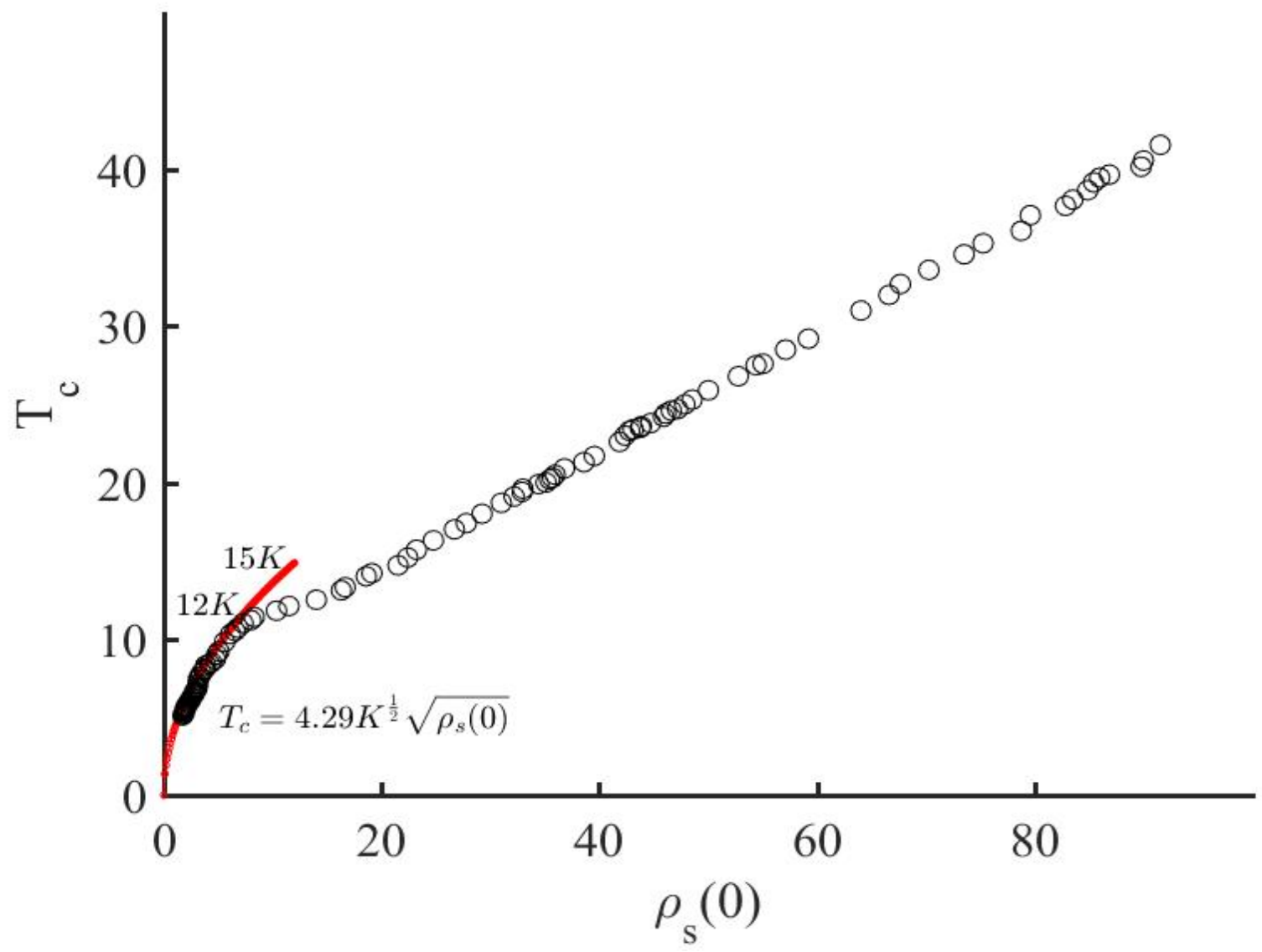

**Figure 1:** The experimental data from [13] are plotted as black circles, which belong to the $T_c$ interval $[5.1\,K,\ 41.6\,K]$. **a.** The theoretical parabolic scaling (red line) $T_c = 4.29\,K^{1/2} \cdot \sqrt{\rho_s(0)}$ perfectly fits the experimental data in $[5.1\,K,\ T_M]$, while the linear scaling (blue line) perfectly fits the experimental data in $[T_Q,\ 41.6\,K]$, where $T_M \approx 11\,K$ and $T_Q \approx 17\,K$, as predicted by equation (36). **b.** The theoretical parabolic scaling (red line) $T_c = 4.29\,K^{1/2} \cdot \sqrt{\rho_s(0)}$ is fitted with the experimental data in the $T_c$ interval $[0, 15\,K]$ , where $T_M \approx 12\,K$ and $T_Q \approx 15\,K$ are experimentally measured [13].

## Acknowledgments

This work was supported by the Fundamental Research Funds for the Central Universities (Grant No. SWU1409444 and Grant No. SWU1809020) and the State Scholarship Fund granted by the China Scholarship Council.